\documentclass[conference]{IEEEtran}

\usepackage[utf8]{inputenc}
\usepackage[T1]{fontenc}
\usepackage{dsfont}

\usepackage[caption=no]{subfig}
\usepackage[linesnumbered]{algorithm2e}
\usepackage{tikz}
\usetikzlibrary{calc,chains,decorations.markings,matrix,patterns,positioning,shapes.geometric}
\usepackage{pgfplots, pgfplotstable}
\usepackage{algorithmicx}
\usepackage{algpseudocode}
\usepackage{array}
\usepackage{tabularx,amsmath,amssymb}
\usepackage{graphicx}
\usepackage{amsmath}
\usepackage{booktabs}
\newcolumntype{Z}{>{\raggedleft\arraybackslash}X}
\newtheorem{definition}{Definition}
\newtheorem{problem}{Problem}
\algdef{SE}[DOWHILE]{Do}{doWhile}{\algorithmicdo}[1]{\algorithmicwhile\ #1}%
\makeatletter
\newcommand{\nosemic}{\renewcommand{\@endalgocfline}{\relax}}
\newcommand{\dosemic}{\renewcommand{\@endalgocfline}{\algocf@endline}}
\makeatother

\SetKwInOut{Input}{Input}
\SetKwInOut{Output}{Output}

\newtheorem{example}{Example}

\usepackage{microtype}

\ifCLASSINFOpdf

\else

\fi

\newcommand{\tocorrect}[1]{\ignorespaces}

\begin{document}

\title{Reversible Pebbling Game for Quantum Memory Management}

\author{\IEEEauthorblockN{Giulia Meuli\IEEEauthorrefmark{1} \quad
Mathias Soeken\IEEEauthorrefmark{1} \quad
Martin Roetteler\IEEEauthorrefmark{2} \quad
Nikolaj Bjorner\IEEEauthorrefmark{2} \quad
Giovanni De Micheli\IEEEauthorrefmark{1}}
\IEEEauthorblockA{\IEEEauthorrefmark{1} EPFL, Lausanne, Switzerland \qquad\IEEEauthorrefmark{2}Microsoft, Redmond, WA, United States}
}

\maketitle

\begin{abstract}
Quantum memory management is becoming a pressing problem, especially given the recent research effort to develop new and more complex quantum algorithms. 
The only existing automatic method for quantum states clean-up relies on the availability of many extra resources. 
In this work, we propose an automatic tool for quantum memory management. We show how this problem exactly matches the reversible pebbling game. Based on that, we develop a SAT-based algorithm that returns a valid clean-up strategy, taking the limitations of the quantum hardware into account. The developed tool empowers the designer with the flexibility required to explore the trade-off between memory resources and number of operations.
We present three show-cases to prove the validity of our approach. First, we apply the algorithm to straight-line programs, widely used in cryptographic applications. Second, we perform a comparison with the existing approach, showing an average improvement of 52.77\%. Finally, we show the advantage of using the tool when synthesizing a quantum circuit on a constrained near-term quantum device.
\end{abstract}

\section{Introduction}
The prospective of experimenting with a practical quantum computer is closing up thanks to the recent developments in hardware technology~\cite{kelly18,knight17,Hempel18}. 
Driven by the revolutionary potential capabilities of quantum computing, research is extremely active both in academic and in industrial environments~\cite{castelvecchi17}.
The race is on to develop quantum algorithms capable of proving quantum supremacy, which is the ability to solve problems that cannot be solved classically~\cite{Shor94, babbush16, reiher17, harrow09}. 

A large part of the design of quantum algorithms is still performed manually, despite the emergence of several automatic methods for both synthesis~\cite{soeken18, vad12} and optimization~\cite{amy19, meuli18, Nam18} of quantum circuits.
Most manual and automatic approaches for quantum circuit synthesis decompose large functionality into smaller parts in order to deal with complexity. Each part requires some resources in terms of qubits and quantum operations. The components can be connected together in order to obtain the desired computation for the overall circuit.

\begin{figure}[t]
  \footnotesize
  \centering
  \subfloat[]{\begin{tikzpicture}[scale=1.000000,x=1pt,y=1pt]
\filldraw[color=white] (0.000000, -6.500000) rectangle (64.000000, 58.500000);
\draw[color=black] (0.000000,52.000000) -- (64.000000,52.000000);
\draw[color=black] (0.000000,52.000000) node[left] {$|x_1\rangle$};
\draw[color=black] (0.000000,39.000000) -- (64.000000,39.000000);
\draw[color=black] (0.000000,39.000000) node[left] {$|x_2\rangle$};
\draw[color=black] (0.000000,26.000000) -- (64.000000,26.000000);
\draw[color=black] (0.000000,26.000000) node[left] {$|0\rangle$};
\draw[color=black] (0.000000,13.000000) -- (64.000000,13.000000);
\draw[color=black] (0.000000,13.000000) node[left] {$|x_3\rangle$};
\draw[color=black] (0.000000,0.000000) -- (64.000000,0.000000);
\draw[color=black] (0.000000,0.000000) node[left] {$|0\rangle$};
\draw (16.000000,52.000000) -- (16.000000,26.000000);
\begin{scope}
\draw[fill=white] (16.000000, 39.000000) +(-45.000000:14.142136pt and 26.870058pt) -- +(45.000000:14.142136pt and 26.870058pt) -- +(135.000000:14.142136pt and 26.870058pt) -- +(225.000000:14.142136pt and 26.870058pt) -- cycle;
\clip (16.000000, 39.000000) +(-45.000000:14.142136pt and 26.870058pt) -- +(45.000000:14.142136pt and 26.870058pt) -- +(135.000000:14.142136pt and 26.870058pt) -- +(225.000000:14.142136pt and 26.870058pt) -- cycle;
\draw (16.000000, 39.000000) node {{$f$}};
\end{scope}
\draw (48.000000,26.000000) -- (48.000000,0.000000);
\begin{scope}
\draw[fill=white] (48.000000, 13.000000) +(-45.000000:14.142136pt and 26.870058pt) -- +(45.000000:14.142136pt and 26.870058pt) -- +(135.000000:14.142136pt and 26.870058pt) -- +(225.000000:14.142136pt and 26.870058pt) -- cycle;
\clip (48.000000, 13.000000) +(-45.000000:14.142136pt and 26.870058pt) -- +(45.000000:14.142136pt and 26.870058pt) -- +(135.000000:14.142136pt and 26.870058pt) -- +(225.000000:14.142136pt and 26.870058pt) -- cycle;
\draw (48.000000, 13.000000) node {{$g$}};
\end{scope}
\draw[color=black] (64.000000,52.000000) node[right] {$|x_1\rangle$};
\draw[color=black] (64.000000,39.000000) node[right] {$|x_2\rangle$};
\draw[color=black] (64.000000,26.000000) node[right] {$?$};
\draw[color=black] (64.000000,13.000000) node[right] {$|x_3\rangle$};
\draw[color=black] (64.000000,0.000000) node[right] {$|y\rangle$};
\end{tikzpicture}}
  \hfil
  \subfloat[]{\begin{tikzpicture}[scale=1.000000,x=1pt,y=1pt]
\filldraw[color=white] (0.000000, -6.500000) rectangle (96.000000, 58.500000);
\draw[color=black] (0.000000,52.000000) -- (96.000000,52.000000);
\draw[color=black] (0.000000,52.000000) node[left] {$|x_1\rangle$};
\draw[color=black] (0.000000,39.000000) -- (96.000000,39.000000);
\draw[color=black] (0.000000,39.000000) node[left] {$|x_2\rangle$};
\draw[color=black] (0.000000,26.000000) -- (96.000000,26.000000);
\draw[color=black] (0.000000,26.000000) node[left] {$|0\rangle$};
\draw[color=black] (0.000000,13.000000) -- (96.000000,13.000000);
\draw[color=black] (0.000000,13.000000) node[left] {$|x_3\rangle$};
\draw[color=black] (0.000000,0.000000) -- (96.000000,0.000000);
\draw[color=black] (0.000000,0.000000) node[left] {$|0\rangle$};
\draw (16.000000,52.000000) -- (16.000000,26.000000);
\begin{scope}
\draw[fill=white] (16.000000, 39.000000) +(-45.000000:14.142136pt and 26.870058pt) -- +(45.000000:14.142136pt and 26.870058pt) -- +(135.000000:14.142136pt and 26.870058pt) -- +(225.000000:14.142136pt and 26.870058pt) -- cycle;
\clip (16.000000, 39.000000) +(-45.000000:14.142136pt and 26.870058pt) -- +(45.000000:14.142136pt and 26.870058pt) -- +(135.000000:14.142136pt and 26.870058pt) -- +(225.000000:14.142136pt and 26.870058pt) -- cycle;
\draw (16.000000, 39.000000) node {{$f$}};
\end{scope}
\draw (48.000000,26.000000) -- (48.000000,0.000000);
\begin{scope}
\draw[fill=white] (48.000000, 13.000000) +(-45.000000:14.142136pt and 26.870058pt) -- +(45.000000:14.142136pt and 26.870058pt) -- +(135.000000:14.142136pt and 26.870058pt) -- +(225.000000:14.142136pt and 26.870058pt) -- cycle;
\clip (48.000000, 13.000000) +(-45.000000:14.142136pt and 26.870058pt) -- +(45.000000:14.142136pt and 26.870058pt) -- +(135.000000:14.142136pt and 26.870058pt) -- +(225.000000:14.142136pt and 26.870058pt) -- cycle;
\draw (48.000000, 13.000000) node {{$g$}};
\end{scope}
\draw (80.000000,52.000000) -- (80.000000,26.000000);
\begin{scope}
\draw[fill=white] (80.000000, 39.000000) +(-45.000000:14.142136pt and 26.870058pt) -- +(45.000000:14.142136pt and 26.870058pt) -- +(135.000000:14.142136pt and 26.870058pt) -- +(225.000000:14.142136pt and 26.870058pt) -- cycle;
\clip (80.000000, 39.000000) +(-45.000000:14.142136pt and 26.870058pt) -- +(45.000000:14.142136pt and 26.870058pt) -- +(135.000000:14.142136pt and 26.870058pt) -- +(225.000000:14.142136pt and 26.870058pt) -- cycle;
\draw (80.000000, 39.000000) node {{$f^{-1}$}};
\end{scope}
\draw[color=black] (96.000000,52.000000) node[right] {$|x_1\rangle$};
\draw[color=black] (96.000000,39.000000) node[right] {$|x_2\rangle$};
\draw[color=black] (96.000000,26.000000) node[right] {$|0\rangle$};
\draw[color=black] (96.000000,13.000000) node[right] {$|x_3\rangle$};
\draw[color=black] (96.000000,0.000000) node[right] {$|y\rangle$};
\end{tikzpicture}}
  \caption{An example of mapping two parts into quantum circuit; (a) does not uncompute the first part, leading to an unknown \emph{garbage} state, (b) does uncompute the first part by computing it again in reverse order.}\label{garbage}
\end{figure}
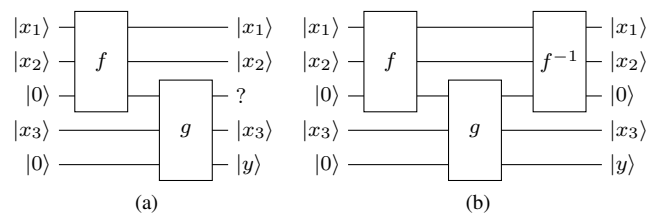

Most of the parts of a large function are used to compute intermediate values, which are stored on qubits. However, the final composed circuit must not emit any of those values. Otherwise, the computed results may entangle with intermediate values and compromise the overall quantum algorithm. Since quantum operations are reversible, intermediate results can be ``uncomputed'' by performing the same operations that computed them, in reverse order.  Fig.~\ref{garbage} illustrates a small example. The composition of the two functions $f$ and $g$ generates an unknown state that can be uncomputed by performing $f$ in reverse order.

There are many possible ways to combine the small parts of a decomposition, each of which resulting in different accumulated costs for number of qubits and number of quantum operations. The requirement that all intermediate results must be uncomputed makes finding a good way to combine parts particularly difficult in quantum computing. Consequently, effective memory management, which guarantees erasure of intermediate results, is crucial in quantum circuit synthesis.

The problem of finding a strategy to compute and uncompute intermediate states for a given fixed number of qubits corresponds to solving the reversible pebbling game. The reversible pebbling game problem has been studied for the first time by Bennett in~\cite{Bennett1989}, in the context of exploring space/time trade-off in reversible computation. Input is a \textit{Directed Acyclic Graph} (DAG), in which each node corresponds to one part of the decomposed computation, and edges define data dependency. Also, nodes can be \emph{pebbled}, meaning that the computed value is available on some resource, in our case on a qubit. The game consists of placing pebbles on the graph nodes.  Initially no node is pebbled.  A pebble can be placed on a node if all its children are pebbled, and the same condition is required to remove a pebble from a node. The game is concluded if all the outputs are pebbled and all the other nodes are unpebbled. Solving the problem returns a valid clean-up strategy.

The problem complexity has been studied in~\cite{Chan2015} where the authors prove that the problem is PSPACE-complete, as the non-reversible pebbling game. An explicit asymptotic expression for the best time-space product is given in~\cite{Knill1995}, while the asymptotic behavior on trees is studied in~\cite{Komarath2018}. 

We propose a solution to the reversible pebbling game that casts the problem as a satisfiability problem. We show how the method is capable of exploring the trade-off between space (number of qubits) and time (number of operations).  In our experimental evaluation, we showcase several examples how our approach can be used to find memory management strategies both for manual and automatic synthesis approaches.

\section{Preliminaries}\label{Pre}

\subsection{Quantum memory management}
Our approach abstracts from the actual quantum operations that are being performed, and therefore the interested reader is referred to the literature for a detailed background on quantum computing~\cite{nielsen00}.

The problem of quantum memory management is crucial in quantum circuit design, as all the garbage states need to be carefully cleaned up. 

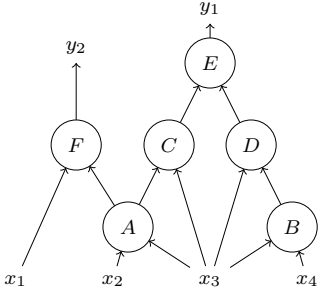
\begin{figure}[t]
  \centering
  \resizebox{!}{4cm}{\tikzset{%
  font=\footnotesize,
  mig/.style={decoration={markings,mark=at position .5 with {\fill (0,0) circle (1.5pt);}}},
  every picture/.style={mig},
  maj/.style={draw,circle,minimum size=15pt,inner sep=0pt,outer sep=0pt},
  small maj/.style={draw,circle,minimum size=10pt,inner sep=0pt,outer sep=0pt},
  big maj/.style={draw,circle,minimum size=20pt,inner sep=0pt,outer sep=0pt},
  computed/.style={fill,pattern=north east lines},
  child/.style={draw,circle,solid,inner sep=0pt,font=\scriptsize},
  title/.style={above,yshift=7pt,text depth=.5ex}
}
\begin{tikzpicture}
  \begin{scope}[every node/.style={big maj},node distance=.4cm]
  	\node (n1){$A$} ;
    \node[right=1.6 of n1] (n2) {$B$};
    
    	\coordinate (c1) at ($(n1.north)!.5!(n2.north)$);    
    \node[above=1.6 of c1] (n4){$E$} ;    
    	\coordinate (c2) at ($(n1.north)!.5!(n4.south)$);
    	\coordinate (c3) at ($(n2.north)!.5!(n4.south)$);
    \node[draw] at (c2) (n3) {$C$};
    \node[draw] at (c3) (n5) {$D$};
    \node[left = .6 of n3] (n6) {$F$};
  \end{scope}
   \begin{scope}
   		\node[above=.8 of n6]  (y2) {$y_2$};
   		\node[above=.2 of n4]  (y1) {$y_1$};
   		\node[below=2.5 of n4.south]  (x3) {$x_3$};
   		\node[left=.8 of x3]  (x2) {$x_2$};
   		\node[left=.8 of x2]  (x1) {$x_1$};
   		\node[right=.8 of x3]  (x4) {$x_4$};
	\end{scope}
	
  \begin{scope}[every node/.style={inner sep=.5pt,below}]
    \draw[<-] (n1) -- (x2);
    \draw[<-] (n1) -- (x3);
    
    \draw[<-] (n2) -- (x4);
    \draw[<-] (n2) -- (x3);
    
    \draw[<-] (n3) -- (n1);
    \draw[<-] (n3) -- (x3);
    
    \draw[<-] (n5) -- (n2);
    \draw[<-] (n5) -- (x3);
    
    \draw[<-] (n4) -- (n3);
    \draw[<-] (n4) -- (n5);
    \draw[->] (n4) -- (y1);
    
    \draw[<-] (n6) -- (x1);
    \draw[<-] (n6) -- (n1);
    \draw[->] (n6) -- (y2);
       
  \end{scope}

\end{tikzpicture}}
  \caption{Example of a DAG}
  \label{DAG}
\end{figure}

\begin{figure}[t]
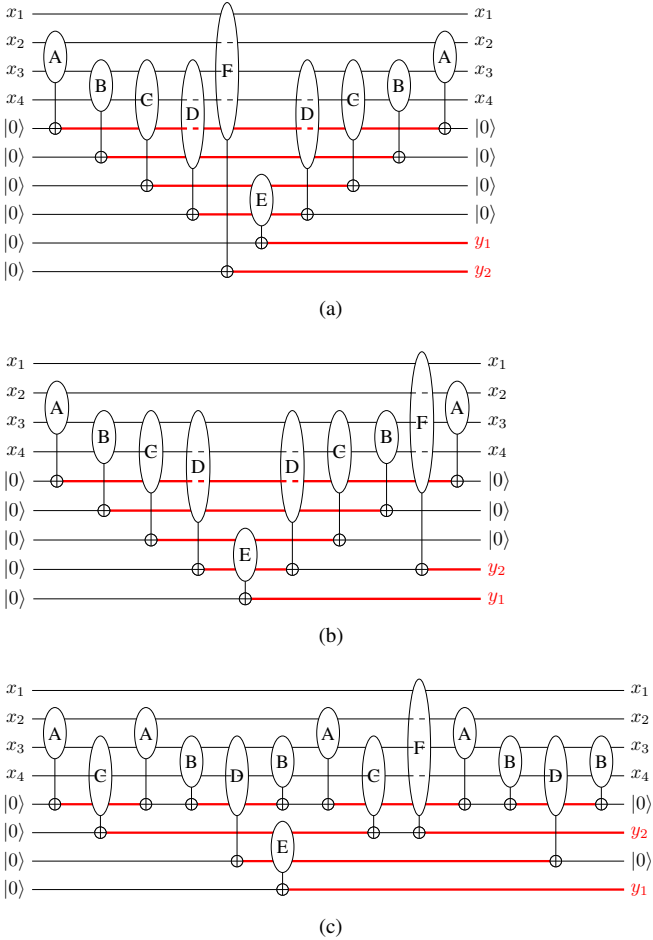

  \centering
  \subfloat[]{\resizebox{\columnwidth}{!}{\input{pictures/stgbennett.tikz}}}
  \newline
  \subfloat[]{\resizebox{\columnwidth}{!}{\input{pictures/stgorder.tikz}}}
  \newline
  \subfloat[]{\resizebox{\columnwidth}{!}{\input{pictures/stgpeb.tikz}}}
  \caption{Three different uncomputing strategies: (a) Bennett strategy; (b) space-optimized by reordering; (c) space-optimized by increasing the number of gates}\label{expeb}
\end{figure}

In the remainder of this section, we consider the example of a quantum algorithm that performs the following mapping: $|x_1\rangle |x_2\rangle |x_3\rangle |x_4\rangle |0\rangle |0\rangle \mapsto |x_1\rangle |x_2\rangle |x_3\rangle |x_4\rangle |y_1\rangle |y_2\rangle$ where
\begin{align*}
z_1 = A(x_2, x_3) \qquad
z_2 = C(z_1, x_3) \qquad
z_3 = B(x_3, x_4) \\
z_4 = D(z_3, x_3) \qquad
y_1 = E(z_2, z_4)	\qquad
y_2 = F(x_1, z_1)
\end{align*} 
with $A, B, C, D, E, F$ being some generic 2-input Boolean operations and $z_1, z_2,z_3, z_4$ the intermediate results.
Such computation can be represented using the DAG in Fig.~\ref{DAG}, in which each node represents an operation. An edge is drawn from a node $v$ to a node $w$, if $w$ requires the value of $v$ (see Fig.~\ref{DAG}).

In order to build the quantum circuit to perform our computation we exploit the direct correspondence between each node in the graph and a reversible single-target gate.
\begin{definition}[Single-target gate]
A single-target gate $G_c$ is a reversible gate characterized by a control function $c$, by a set of control qubits $q_1,\dots,q_k$ and by a target qubit $q_t$. The gate inverts the value of the target line if $c(q_1, \dots, q_k)$ evaluates to true, i.e., 
\[G_c : |q_1\rangle \dots |q_k\rangle|q_t\rangle \mapsto  |q_1\rangle \dots |q_k\rangle|q_t \oplus c(q_1,\dots,q_k)\rangle\]

\end{definition}
\smallskip\noindent Different reversible circuits resulting from this translation is shown in Fig.~\ref{expeb}.
When two identical gates are performed twice on the same target, the value on the line is uncomputed, and goes back to its original state.
Qubits initialized to $|0\rangle$, called \textit{ancillae}, are used to store the intermediate results and must be restored after the computation.
Once the results $y_1$ and $y_2$ have been computed, all the intermediate values $z_1,z_2,z_3,z_4$ must be cleaned up. 

A simple solution is the one proposed in Fig.~\ref{expeb}(a), 
which is referred to as the Bennett~\cite{Bennett1989} strategy. It consists of computing all the operations in a bottom-up order, and then uncomputing the intermediate results in a top-down fashion, so that all the nodes have their inputs available. This strategy always leads to the minimum number of gates, and to the maximum number of qubits.
The order in which the DAG is converted into a reversible circuit can have a significant effect on how the memory is managed. In the example strategy in Fig.~\ref{expeb}(b) it is shown how, only by changing the order of the operations, it is possible to save one qubit, without increasing the duration of the computation.
Finally, by allowing an increase in the number of gates, we can further reduce the number of ancillae to 4. In this case some functions are computed several times, see Fig.~\ref{expeb}(c).

In Fig.~\ref{expeb} ancillae are colored red during the time they are storing an intermediate result.  The first two strategies store values for a long time during which they are not needed, whereas the last strategy makes a good usage of fewer memory. Those three graphs are useful to visualize the trade-off between hardware resources (i.e., qubits) and time (i.e., operations). 
We cannot state that the last method performs better than the first one, as that would depend from the actual hardware constraints. What we achieve in this work is to empower the designer with the ability to choose whether exchange memory for time and vice versa.

\subsection{Reversible pebbling game}\label{rpg}
The problem of finding the best uncomputing strategy is equivalent to the reversible pebbling game problem. In the remainder we use independently \textit{pebbling} and \textit{uncomputing} strategy.

\begin{definition}[Reversible pebbling configuration]
A reversible pebbling configuration of a DAG $G = (V, E)$ is the set $P\subseteq V$ of all the pebbled vertices.
\end{definition}

\begin{definition}[Reversible pebbling strategy]
A reversible pebbling strategy of a DAG $G$ is a sequence of reversible pebbling configurations 
$P = (P_1,...,P_m)$ such that $P_1 = \{ \}$ and $P_m = O$, where $O$ is the set of all sinks of $G$.
For each $1 < i\leq m$, we have $P_i = P_{i-1} \cup \{v\}$ or $P_i = P_{i-1} / \{v\}$ and $P_i \neq P_{i-1}$. 
All in-neighbors of $v$ are in $P_{i-1}$.
\end{definition}

\begin{figure}[t] 
  \centering
  \resizebox{!}{1.7cm}{\begin{tikzpicture}
[box/.style={rectangle,draw=black,thick, minimum size=1cm},]
\node[box] at (0, 5){};
\node[] at (-1, 5){\Huge{F}};
\node[box] at (0, 4){};
\node[] at (-1, 4){\Huge{E}};
\node[box] at (0, 3){};
\node[] at (-1, 3){\Huge{D}};
\node[box] at (0, 2){};
\node[] at (-1, 2){\Huge{C}};
\node[box] at (0, 1){};
\node[] at (-1, 1){\Huge{B}};
\node[box] at (0, 0){};
\node[] at (-1, 0){\Huge{A}};
\node[box] at (1, 5){};
\node[box] at (1, 4){};
\node[box] at (1, 3){};
\node[box] at (1, 2){};
\node[box] at (1, 1){};
\node[box, fill=black] at (1, 0){};
\node[box] at (2, 5){};
\node[box] at (2, 4){};
\node[box] at (2, 3){};
\node[box] at (2, 2){};
\node[box, fill=black] at (2, 1){};
\node[box, fill=black] at (2, 0){};
\node[box] at (3, 5){};
\node[box] at (3, 4){};
\node[box] at (3, 3){};
\node[box, fill=black] at (3, 2){};
\node[box, fill=black] at (3, 1){};
\node[box, fill=black] at (3, 0){};
\node[box] at (4, 5){};
\node[box] at (4, 4){};
\node[box, fill=black] at (4, 3){};
\node[box, fill=black] at (4, 2){};
\node[box, fill=black] at (4, 1){};
\node[box, fill=black] at (4, 0){};
\node[box] at (5, 5){};
\node[box, fill=black] at (5, 4){};
\node[box, fill=black] at (5, 3){};
\node[box, fill=black] at (5, 2){};
\node[box, fill=black] at (5, 1){};
\node[box, fill=black] at (5, 0){};

\node[box, fill=black] at (6, 5){};
\node[box, fill=black] at (6, 4){};
\node[box, fill=black] at (6, 3){};
\node[box, fill=black] at (6, 2){};
\node[box, fill=black] at (6, 1){};
\node[box, fill=black] at (6, 0){};
\node[box, fill=black] at (7, 5){};
\node[box, fill=black] at (7, 4){};
\node[box] at (7, 3){};
\node[box, fill=black] at (7, 2){};
\node[box, fill=black] at (7, 1){};
\node[box, fill=black] at (7, 0){};
\node[box, fill=black] at (8, 5){};
\node[box, fill=black] at (8, 4){};
\node[box] at (8, 3){};
\node[box] at (8, 2){};
\node[box, fill=black] at (8, 1){};
\node[box, fill=black] at (8, 0){};
\node[box, fill=black] at (9, 5){};
\node[box, fill=black] at (9, 4){};
\node[box] at (9, 3){};
\node[box] at (9, 2){};
\node[box] at (9, 1){};
\node[box, fill=black] at (9, 0){};
\node[box, fill=black] at (10, 5){};
\node[box, fill=black] at (10, 4){};
\node[box] at (10, 3){};
\node[box] at (10, 2){};
\node[box] at (10, 1){};
\node[box] at (10, 0){};

\end{tikzpicture}}    
  \resizebox{!}{1.7cm}{\begin{tikzpicture}
[box/.style={rectangle,draw=black,thick, minimum size=1cm},]
\node[] at (-1, 5){\Huge{F}};
\node[] at (-1, 4){\Huge{E}};
\node[] at (-1, 3){\Huge{D}};
\node[] at (-1, 2){\Huge{C}};
\node[] at (-1, 1){\Huge{B}};
\node[] at (-1, 0){\Huge{A}};
\node[box] at (0, 0){};
\node[box] at (0, 1){};
\node[box] at (0, 2){};
\node[box] at (0, 3){};
\node[box] at (0, 4){};
\node[box] at (0, 5){};
\node[box, fill=black] at (1, 0){};
\node[box] at (1, 1){};
\node[box] at (1, 2){};
\node[box] at (1, 3){};
\node[box] at (1, 4){};
\node[box] at (1, 5){};
\node[box, fill=black] at (2, 0){};
\node[box] at (2, 1){};
\node[box, fill=black] at (2, 2){};
\node[box] at (2, 3){};
\node[box] at (2, 4){};
\node[box] at (2, 5){};
\node[box] at (3, 0){};
\node[box] at (3, 1){};
\node[box, fill=black] at (3, 2){};
\node[box] at (3, 3){};
\node[box] at (3, 4){};
\node[box] at (3, 5){};
\node[box] at (4, 0){};
\node[box, fill=black] at (4, 1){};
\node[box, fill=black] at (4, 2){};
\node[box] at (4, 3){};
\node[box] at (4, 4){};
\node[box] at (4, 5){};
\node[box] at (5, 0){};
\node[box, fill=black] at (5, 1){};
\node[box, fill=black] at (5, 2){};
\node[box, fill=black] at (5, 3){};
\node[box] at (5, 4){};
\node[box] at (5, 5){};

\node[box] at (6, 0){};
\node[box] at (6, 1){};
\node[box, fill=black] at (6, 2){};
\node[box, fill=black] at (6, 3){};
\node[box] at (6, 4){};
\node[box] at (6, 5){};

\node[box] at (7, 0){};
\node[box] at (7, 1){};
\node[box, fill=black] at (7, 2){};
\node[box, fill=black] at (7, 3){};
\node[box, fill=black] at (7, 4){};
\node[box] at (7, 5){};

\node[box, fill=black] at (8, 0){};
\node[box] at (8, 1){};
\node[box, fill=black] at (8, 2){};
\node[box, fill=black] at (8, 3){};
\node[box, fill=black] at (8, 4){};
\node[box] at (8, 5){};

\node[box, fill=black] at (9, 0){};
\node[box] at (9, 1){};
\node[box] at (9, 2){};
\node[box, fill=black] at (9, 3){};
\node[box, fill=black] at (9, 4){};
\node[box] at (9, 5){};

\node[box, fill=black] at (10, 0){};
\node[box] at (10, 1){};
\node[box] at (10, 2){};
\node[box, fill=black] at (10, 3){};
\node[box, fill=black] at (10, 4){};
\node[box, fill=black] at (10, 5){};

\node[box] at (11, 0){};
\node[box] at (11, 1){};
\node[box] at (11, 2){};
\node[box, fill=black] at (11, 3){};
\node[box, fill=black] at (11, 4){};
\node[box, fill=black] at (11, 5){};

\node[box] at (12, 0){};
\node[box, fill=black] at (12, 1){};
\node[box] at (12, 2){};
\node[box, fill=black] at (12, 3){};
\node[box, fill=black] at (12, 4){};
\node[box, fill=black] at (12, 5){};

\node[box] at (13, 0){};
\node[box, fill=black] at (13, 1){};
\node[box] at (13, 2){};
\node[box] at (13, 3){};
\node[box, fill=black] at (13, 4){};
\node[box, fill=black] at (13, 5){};

\node[box] at (14, 0){};
\node[box] at (14, 1){};
\node[box] at (14, 2){};
\node[box] at (14, 3){};
\node[box, fill=black] at (14, 4){};
\node[box, fill=black] at (14, 5){};

\end{tikzpicture}}
  \caption{Two different pebbling strategies for the DAG in the example.}
  \label{3}
\end{figure}
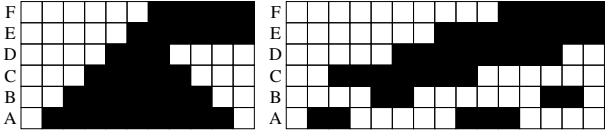

In Fig.~\ref{3} we show two possible pebbling strategies for the DAG in the example. Each row of the grids represents a node and each column, from left to right, corresponds to a single step. A black square means that the corresponding node is pebbled. 
In accordance to the rules of the game, the initial configuration is empty and the last only contains output vertices. In these examples we only allow one move per step. The first strategy is the one reported by Bennett~\cite{Bennett1989}, in which we naively compute all the nodes and then uncompute intermediate results. This pebbling requires a number of pebbles equal to the number of nodes, 6 in the example, and only 10 steps, that is minimum. The complete sequence of pebbling configurations for the first example is:
{\small
\begin{align*}
P_0&=\{ \} & P_6&=\{A, B, C, D, E, F\} \\
P_1&=\{A\} & P_7&=\{A, B, C, E, F\} \\
P_2&=\{A, B\} & P_8&=\{A, B, E, F\} \\
P_3&=\{A, B, C \} & P_9&=\{A, E, F\} \\
P_4&=\{A, B, C, D\} & P_{10}&=\{E, F\} \\
P_5&=\{A, B, C, D, E\} 
\end{align*}
}%
The second approach is a strategy that only uses 4 pebbles. To reduce the number of pebbles, it computes twice the nodes $a$ and $b$, increasing the number of steps to 14. The complete sequence of pebbling configurations for the second example is:
{\small
\begin{align*}
P_0&=\{ \} & P_8&=\{A, C, D, E\} \\
P_1&=\{A\} & P_9&=\{A, D, E\} \\
P_2&=\{A, C\} & P_{10}&=\{A, D, E, F\} \\
P_3&=\{ C \} & P_{11}&=\{D, E, F\} \\
P_4&=\{ B, C\} & P_{12}&=\{B, D, E, F\} \\
P_5&=\{ B, C, D\} & P_{13}&=\{B, E, F\} \\
P_6&=\{ C, D\} & P_{14}&=\{E, F\} \\
P_7&=\{ C, D, E\}
\end{align*} 
}%

\subsection{SAT solver}
Given a Boolean function $f(x_1,...,x_n)$, the Boolean satisfiability problem (or SAT problem) consists of determining an assignment $V$ to the variables $x_1, ... , x_n$ such that $f$ is satisfied (evaluates to true). If such an assignment exists we call it a satisfying assignment, otherwise the problem is unsatisfiable.
SAT solvers are software programs in which $f$ is specified as conjunctive normal form (CNF) consisting of a conjunction of clauses where each clause is a disjunction of literals. We define a literal as instance of a variable or its complement. SAT can be summarized as follows: given a list of clauses, determine if there exists a variable assignment that satisfies all of them.

\section{SAT-encoding}
In this work we aim at finding a good pebbling strategy while constraining the maximum number of pebbles per step.
\begin{problem}
Given a DAG and a number of pebbles, find a valid pebbling strategy using the minimum number of steps.
\end{problem}
As we use a SAT solver~\cite{Nik08} to extract our solution, we have to decompose this problem into many SAT problems. 
\begin{problem} \label{p2}
Given a DAG and $K$ pebbles, does a valid pebbling strategy with $K$ steps exist? 
\end{problem}
The solver can either find a solution and return a pebbling strategy, or state that no solution exists. In this case we increase the number of steps to $K+1$ until a satisfying solution is found.

Following the definition of a reversible pebbling game given in Section~\ref{rpg}, we first declare our set of variables, and then we impose satisfiability constraints.

\subsection{Variables}
The input DAG $G=(V,E)$ has some nodes which compute an output value and we refer to them as a set $O \subseteq V$.  Note that the primary inputs are \emph{not} nodes in the DAG.  We also define $C(v) = \{w \mid w \rightarrow v\}$ as all children of a node $v$.
\begin{example}
The DAG in Fig.~\ref{DAG} has six nodes $\{A, B, C, D, E, F\}$ and two outputs $O = \{E, F\}$.  Note that, e.g., $C(A) = \emptyset$, since the primary inputs are not part of the DAG.
\end{example}
Problem~\ref{p2} is encoded in terms of the 
\textit{pebble state variables $p_{v,i}$}. For $v \in V$ and $0 \le i \le K$, those are Boolean variables that evaluate to true if the node $v$ is pebbled at time $i$.
Note that the SAT formula encodes $K + 1$ pebble configurations with $K$ steps describing the transition from one configuration to the other.
\subsection{Clauses}
The following set of clauses describes the reversible pebbling problem.
\begin{itemize}
\item\textit{Initial and final clauses:} at time $0$ all the nodes are unpebbled and at time $K$ all the outputs need to be pebbled and all the intermediate results unpebbled
\[
\bigwedge\limits_{v \in V} \bar p_{v,0}  \wedge  \bigwedge\limits_{v \in O} p_{v,K} \wedge  \bigwedge\limits_{v \not\in O} \bar p_{v,K} 
\]

\item\textit{Move clauses:} if a node is pebbled or unpebbled at time $i+1$, then all its children are pebbled at time $i$ and time $i+1$
\[
\bigwedge\limits_{i=1}^{K} \bigwedge \limits_{(v, w) \in E} ((p_{v, i} \oplus p_{v, i+1}) \rightarrow (p_{w, i} \wedge p_{w, i+1}))
\]

\item\textit{Cardinality clauses:} at each step, at most $P$ pebbles are used
\[
\bigwedge\limits_{i=0}^{K} (\sum\limits_{v\in V} p_{v, i} \leq P)
\]
\end{itemize}
\begin{table*}[t!]
\centering
\caption{Comparison}
\begin{tabular}{lrrrrrrrrrr}
\toprule
\multicolumn{4}{l}{}&\multicolumn{2}{c}{Bennett}&\multicolumn{3}{c}{Pebbling strategy}& &  \\
    &  pi  &  po  &  nodes  &  P  &  K  &  P  &  K  &  runtime  &  \%P  &  $\times$K  \\ 
\midrule
b2\_m3 &8	&8	&74	    &66	    &124	 &30	    &186	&0.17	&54.55	&1.5 \\
b3\_m4 &12	&12	&59	    &47	    &82	    &20	    &117	&121.37	&57.45	&1.43\\
b4\_m5 &16	&16	&203	&187	&358	&83	    &778	&55.75	&55.61	&2.17\\
b5\_m7 &20	&20	&256	&236	&452	&106	&888	&31.15	&55.08	&1.96\\
b6\_m7 &24	&24	&310	&286	&548	&130	&1132	&35.72	&54.55	&2.07\\
b8\_m7 &32	&32	&422	&390	&748	&187	&1884	&11.59	&52.05	&2.52\\
b10\_m7 &40	&40	&535	&495	&950	&264	&2938	&28.66	&46.67	&3.09\\
b12\_m7 &48	&48	&646	&598	&1148	&331	&4228	&56.33	&44.65	&3.68\\
b16\_m23 &64	&64	&881	&817	&1570	&480	&6218	&133.45	&41.25	&3.96\\
\midrule
c17	&5	&2	&12	&7	&12	&4	&12	&0.01	&42.86	&1\\
c432	&36	&7	&208	&172	&337	&60	&685	&23.70	&65.12	&2.03\\
c499	&41	&32	&219	&178	&324	&77	&610	&60.08	&56.74	&1.88\\
c880	&60	&26	&334	&274	&522	&82	&1280	&43.52	 &70.07	&2.45\\
c1355	&41	&32	&219	&178	&324	&77	&594	&2.63	&56.74	&1.83\\
c1908	&33	&25	&220	&187	&349	&70	&875	&57.97	    &62.57	&2.51\\
c2670	&157	&63	&554	&397	&731	&160	&1948	&47.94	&59.7	&2.66\\
c3540	&50	&22	&856	&806	&1590	&416	&5434	&111.20	&48.39	&3.42\\
c5315	&178	&123	&1257	&1079	&2035	&498	&7635	&118.38	&53.85	&3.75\\
c6288	&32	&32	&1011	&979	&1926	&640	&10232	&101.31	&34.63	&5.31\\
c7552	&207	&108	&1151	&944	&1780	&540	&7757	&124.1	&42.8	 &4.36\\
\midrule
\multicolumn{11}{l}{Average percentage reduction of pebbles = 52.77}\\
\multicolumn{11}{l}{Average multiplicative factor for the number of steps = 2.68}\\
\bottomrule
\end{tabular}

\label{tab}
\end{table*}
\begin{figure*}[h]
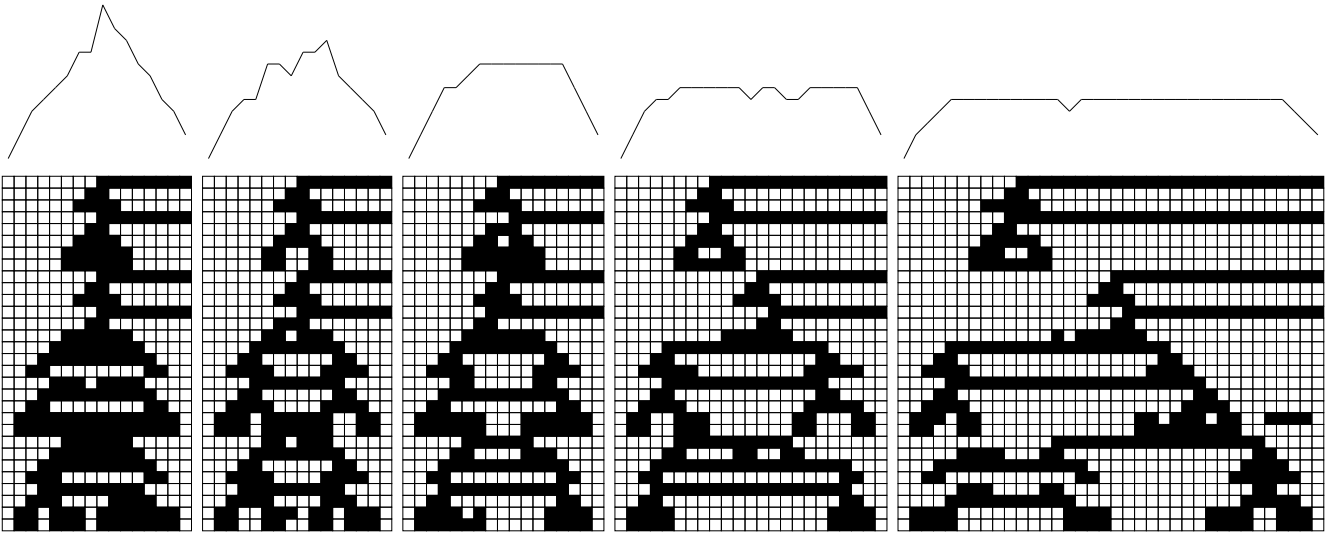

  \centering
  \resizebox{!}{7cm}{\input{pictures/pstrat.tikz}}
  \resizebox{!}{7cm}{\input{pictures/pstrat20.tikz}}
  \resizebox{!}{7cm}{\input{pictures/pstrat16.tikz}}
  \resizebox{!}{7cm}{\input{pictures/pstrat12.tikz}}
  \resizebox{!}{7cm}{\input{pictures/pstrat10.tikz}}
 
  \caption{Example of how the tool can be used to map a computation into a given number of ancillae: respectively 24 (Add:28, Sub:20, Sqrt:15, Mult:11), 20 (Add:36, Sub:32, Sqrt:21, Mult:9), 16 (Add:28, Sub:24, Sqrt:17, Mult:13) , 12 (Add:24, Sub:34, Sqrt:19, Mult:13) and 10 (Add:34, Sub:38, Sqrt:25, Mult:13).}
 \label{Doubling}
\end{figure*}

\begin{figure*}[t]
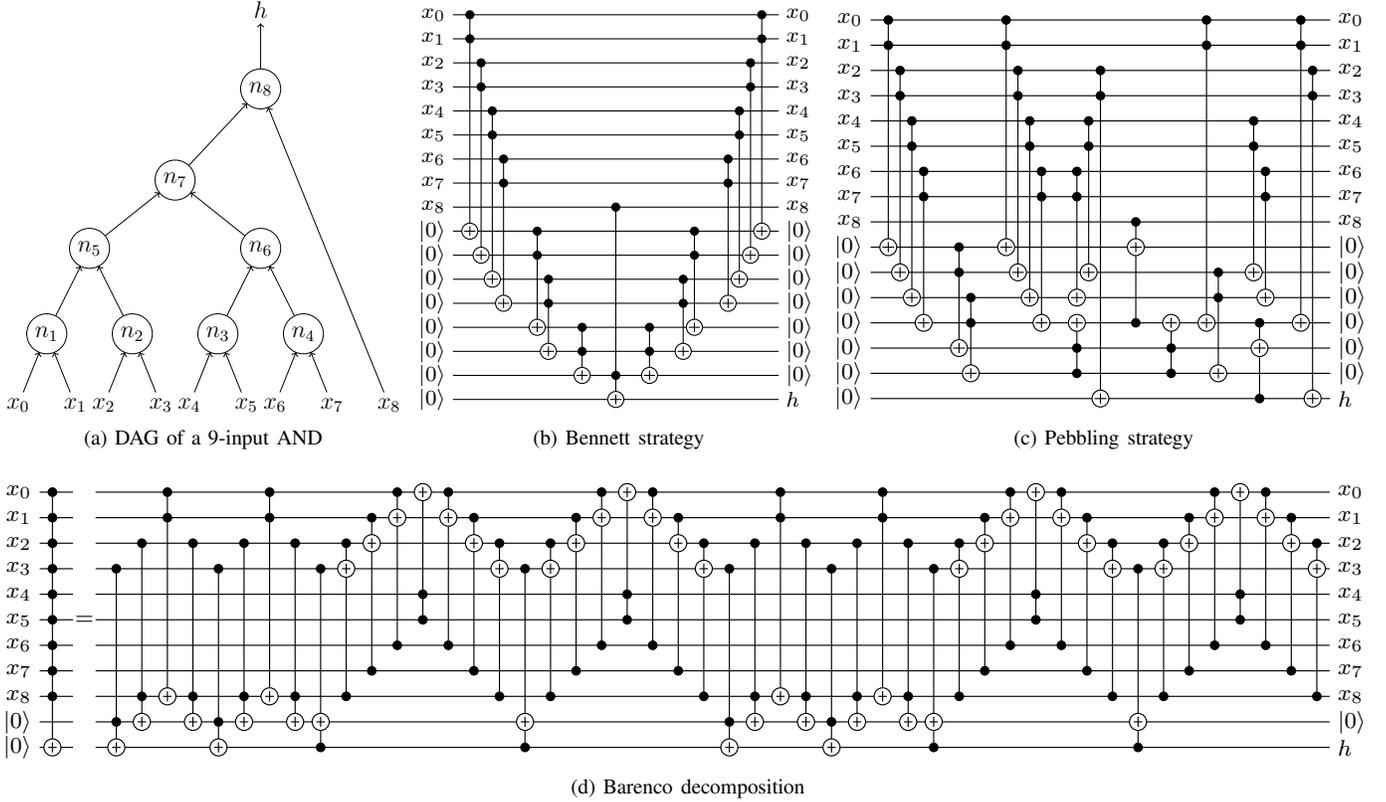

  \footnotesize
  \centering
   \subfloat[DAG of a 9-input AND]{\resizebox{.3\textwidth}{!}{\tikzset{%
  font=\footnotesize,
  mig/.style={decoration={markings,mark=at position .5 with {\fill (0,0) circle (1.5pt);}}},
  every picture/.style={mig},
  maj/.style={draw,circle,minimum size=15pt,inner sep=0pt,outer sep=0pt},
  small maj/.style={draw,circle,minimum size=10pt,inner sep=0pt,outer sep=0pt},
  big maj/.style={draw,circle,minimum size=20pt,inner sep=0pt,outer sep=0pt},
  computed/.style={fill,pattern=north east lines},
  child/.style={draw,circle,solid,inner sep=0pt,font=\scriptsize},
  title/.style={above,yshift=7pt,text depth=.5ex}
}
\begin{tikzpicture}
  \begin{scope}[every node/.style={big maj},node distance=.4cm]
  	\node (n1){\large{$n_1$}} ;
    \node[right=.8 of n1] (n2) {\large{$n_2$}};
    \node[right=.8 of n2] (n3) {\large{$n_3$}};
    \node[right=.8 of n3] (n4) {\large{$n_4$}};
    
    	\coordinate (c1) at ($(n1.north)!.5!(n4.north)$);   
    	\coordinate (c2) at ($(n1.north)!.5!(n2.north)$);   
    	\coordinate (c3) at ($(n3.north)!.5!(n4.north)$);    
    \node[above=2 of c1] (n7){\large{$n_7$}} ;    
    \node[above=.8 of c2] (n5){\large{$n_5$}} ; 
    \node[above=.8 of c3] (n6){\large{$n_6$}} ; 
    \node[above=2.1 of n6](n8){\large{$n_8$}} ;
    
  \end{scope}
   \begin{scope}
   		\node[below=.8 of n1]  (k) {};
   		\node[left=.1 of k]  (x0) {\large{$x_0$}};
   		\node[right=.1 of k]  (x1) {\large{$x_1$}};
   		
   		\node[below=.8 of n2]  (j) {};
   		\node[left=.1 of j]  (x2) {\large{$x_2$}};
   		\node[right=.1 of j]  (x3) {\large{$x_3$}};
   		
   		\node[below=.8 of n3]  (p) {};
   		\node[left=.1 of p]  (x4) {\large{$x_4$}};
   		\node[right=.1 of p]  (x5) {\large{$x_5$}};
   		
   		\node[below=.8 of n4]  (q) {};
   		\node[left=.1 of q]  (x6) {\large{$x_6$}};
   		\node[right=.1 of q]  (x7) {\large{$x_7$}};
   		
   		\node[above=.8 of n8]  (h) {\large{$h$}};
   		\node[right=.4 of x7]  (x8) {\large{$x_8$}};
	\end{scope}
	
  \begin{scope}[every node/.style={inner sep=.5pt,below}]
    \draw[<-] (n1) -- (x0);
    \draw[<-] (n1) -- (x1);
    
     \draw[<-] (n2) -- (x2);
    \draw[<-] (n2) -- (x3);
    
     \draw[<-] (n3) -- (x4);
    \draw[<-] (n3) -- (x5);
    
     \draw[<-] (n4) -- (x6);
    \draw[<-] (n4) -- (x7);
    
     \draw[<-] (n5) -- (n1);
    \draw[<-] (n5) -- (n2);
    
    \draw[<-] (n6) -- (n3);
    \draw[<-] (n6) -- (n4);
    
    \draw[<-] (n7) -- (n5);
    \draw[<-] (n7) -- (n6);
    
    \draw[<-] (n8) -- (n7);
    \draw[->] (n8) -- (h);
    \draw[<-](n8) -- (x8);

  \end{scope}

\end{tikzpicture}}}
  \subfloat[Bennett strategy]{\resizebox{.3\textwidth}{!}{\input{pictures/bennett.tikz}}}
  \subfloat[Pebbling strategy]{\resizebox{.4\textwidth}{!}{\input{pictures/pebbling.tikz}}}

  \hfill
  \subfloat[Barenco decomposition]{\resizebox{\textwidth}{!}{\input{pictures/barenco.tikz}}}
  \caption{Show-case: map the function $AND(x_0,\dots, x_8)$ into a 16-qubits quantum computer. }\label{and}
\end{figure*}

\section{Applications and Results}
In this section we illustrate the validity of our proposed approach by show-casing several examples in which large computations are expressed in terms of a sequence of smaller ones.  In order to optimally exploit qubit resources, a high-quality quantum memory management is required that can be addressed using our SAT-based reversible pebble game solver. Our algorithm uses at its core the open source SAT solver Z3~\cite{Nik08}.

\subsection{Show-case: Straight-line programs}
We apply our method to the synthesis of straight-line programs used in cryptographic applications. Those programs are combinations of modular arithmetic operations as addition, subtraction, multiplication, and squaring. We assume that for each operation a quantum implementation exists, and will have a given cost in terms of quantum gates and ancillae. We can estimate the cost of an algorithm implementation in terms of number of different operations, according to the resources available.
We choose a straight-line program that implements the addition between two points of an Edward curve in projective coordinates from~\cite{Costello2013}. We pebble the resulting DAG using different number of pebbles. Fig.~\ref{Doubling} visualizes the pebbling strategies obtained with 24, 20, 16, 12, and 10 pebbles. In each case, we obtain a different number of operations, as reported in Fig.~\ref{Doubling}. For example the first implementation performs a total of 74 operations: 28 additions, 20 subtractions, 15 squaring and 11 multiplication. We can see how the tool manages to fit the desired computation into limited number of qubits, by increasing the number of required steps. As a consequence, the last implementation has an higher cost in terms of operations: 110 in total. The overall cost of the algorithm on different hardware can be evaluated having some estimates of the real cost of each operation.
On the top of each grid, we show the dynamic change in the memory employed during the computation. A flat dynamic suggests that a constant number of qubits is used through the whole computation. A solution with a lower peak requires less qubits.

\subsection{Show-case: Comparison with Bennett strategy}
The second show-case has the purpose of quantifying the ability of the program to map a design in a limited number of qubits. We consider an operator called $H$ (different from the Hadamard gate) that is used internally to the algorithm that computes the doubling of two points referred before~\cite{Costello2013}. This operator is a composition of modular additions ($+$) and modular subtraction ($-$);
it has $a, b, c, d$ as inputs and four outputs $x, y, z, t$, where:
\begin{align*}
&t_1= a + b\quad
t_2= c+d\quad
t_3= a - b\quad
t_4= c - d \\
&x=t_1 + t_2\quad
y= t_1 - t_2\quad
z= t_3 + t_4\quad
t= t_3 -  t_4 
\end{align*}
Experiments reported in Table~\ref{tab} show a comparison between the Bennett pebbling method and our solution. The different designs correspond to the $H$ operator with different bitwidths and modulus. We also show the results for the well known ISCAS benchmark. The graph representation for the function has been extracted from an XOR-majority graph using the open source tool \emph{mockturtle}~\cite{SRH18}. A method to use XOR-majority graphs into quantum circuits using a naive quantum memory management strategy was presented in~\cite{sm17}. The number of pebbles corresponds to the minimum one for which the solver could find a solution within 2 minutes. Even with this short timeout, the algorithm finds a solution for a significantly reduced number of pebbles. The average percentage reduction is $52.77\%$. As the pebbles are reduced, the number of steps increase with respect to the Bennett method. In average, the number of steps for the constrained design is $2.68\times$ the one of the naive strategy. With the increase of the size of the DAG, we see a fewer pebble reduction. The reason is in the timeout chosen, as the solver requires more time to deal with large designs: the number of variables of the SAT problem is proportional to $n^2$ where $n$ is the number of nodes of the DAG. Also increasing the number of nodes, more steps seems to be required. This is also dependent on the timeout. In fact the algorithm is capable of finding many solutions with different number of pebbles but same number of steps. Nevertheless more constrained solutions require more time to be resolved.

\subsection{Show-case: Adapting to hardware constraints}
In the last example, we consider programming a quantum system composed of 16 qubits, as for example the ibmqx5 quantum computer from IBM~\cite{ibm17}. We want to map an oracle for the 9-input AND function as part of a more complex algorithm, e.g., Grover's algorithm~\cite{grover96}. The function can be represented using the DAG in Fig.~\ref{and}(a).

Our first attempt is to use the Bennett strategy for qubits clean-up. This method leads to the circuit shown in Fig.~\ref{and}(b). This 17-qubits design cannot be mapped on the chosen 16-bits hardware. 
A second possibility is to apply the well known decomposition method proposed by Barenco in~\cite{barenco95} to the 9-controls Toffoli gates, as in Fig.~\ref{and}(d). In this case only one extra ancilla is required (11 qubits in total). The drawback of this solution is that there is an explosion in the number of gates: from 15 to 48. It is known how, increasing the number of gates can negatively affect the noise in the final result~\cite{linke17}.
Our tool provides with enough flexibility to find a more balanced solution. Setting the number of pebbles to exactly 16 we indeed obtain a valid pebbling strategy that maps the desired functionality into the available number of qubits. The result is a circuit with only 23 gates shown in Fig.~\ref{and}(c).

\section{Conclusion}
We developed a SAT-based algorithm for quantum memory management. We show how the clean-up problem corresponds to the reversible pebbling game problem. Consequently, our algorithm solves instances of the reversible pebbling game to explore the trade-off between memory and number of operations. Finding an efficient pebbling strategy is crucial in quantum algorithm development, where often small manually optimized circuits are cascaded together. Our tool can enable computations in constrained systems, when this would not be possible using the strategies in the literature. 
We show three different show-cases to demonstrate the efficiency of our method. In general, it can be used in cryptographic applications to synthesize straight-line programs, but also in any hierarchical synthesis automatic method. It can be used to estimate the cost of performing an algorithm on a given hardware in terms of number of operations. Our experiments show that we are capable of finding solutions with an average reduction in number of ancillae required of 52.77\% with a timeout of 2 minutes. Finally we give a simple example of how the tool could be used by a designer to map a required computation into an available constrained hardware.  

{\small
\subsubsection*{Acknowledgments}  This research was
supported by H2020-ERC-2014-ADG 669354 CyberCare and the Swiss
National Science Foundation (200021-169084 MAJesty and and
200021-146600).
}%

\bibliographystyle{IEEEtran}

\begin{thebibliography}{10}
\providecommand{\url}[1]{#1}
\csname url@samestyle\endcsname
\providecommand{\newblock}{\relax}
\providecommand{\bibinfo}[2]{#2}
\providecommand{\BIBentrySTDinterwordspacing}{\spaceskip=0pt\relax}
\providecommand{\BIBentryALTinterwordstretchfactor}{4}
\providecommand{\BIBentryALTinterwordspacing}{\spaceskip=\fontdimen2\font plus
\BIBentryALTinterwordstretchfactor\fontdimen3\font minus
  \fontdimen4\font\relax}
\providecommand{\BIBforeignlanguage}[2]{{%
\expandafter\ifx\csname l@#1\endcsname\relax
\typeout{** WARNING: IEEEtran.bst: No hyphenation pattern has been}%
\typeout{** loaded for the language `#1'. Using the pattern for}%
\typeout{** the default language instead.}%
\else
\language=\csname l@#1\endcsname
\fi
#2}}
\providecommand{\BIBdecl}{\relax}
\BIBdecl

\bibitem{kelly18}
J.~Kelly, ``A preview of bristlecone, google’s new quantum processor,''
  \emph{Google Research Blog}, vol.~5, 2018.

\bibitem{knight17}
W.~Knight, ``{IBM} raises the bar with a 50-qubit quantum computer,''
  \emph{Sighted at MIT Review Technology}, 2017.

\bibitem{Hempel18}
C.~e.~A. Hempel, ``Quantum chemistry calculations on a trapped-ion quantum
  simulator,'' \emph{Phys. Rev. X}, 2018.

\bibitem{castelvecchi17}
D.~Castelvecchi, ``Quantum computers ready to leap out of the lab in 2017,''
  \emph{Nature News}, vol. 541, no. 7635, p.~9, 2017.

\bibitem{Shor94}
P.~W. Shor, ``Algorithms for quantum computation: discrete logarithms and
  factoring,'' in \emph{Proceedings 35th Annual Symposium on Foundations of
  Computer Science}, Nov 1994, pp. 124--134.

\bibitem{babbush16}
R.~Babbush, D.~W. Berry, I.~D. Kivlichan, A.~Y. Wei, P.~J. Love, and
  A.~Aspuru-Guzik, ``Exponentially more precise quantum simulation of fermions
  in second quantization,'' \emph{New Journal of Physics}, vol.~18, no.~3, p.
  033032, 2016.

\bibitem{reiher17}
M.~Reiher, N.~Wiebe, K.~M. Svore, D.~Wecker, and M.~Troyer, ``Elucidating
  reaction mechanisms on quantum computers,'' \emph{Proceedings of the National
  Academy of Sciences}, p. 201619152, 2017.

\bibitem{harrow09}
A.~W. Harrow, A.~Hassidim, and S.~Lloyd, ``Quantum algorithm for linear systems
  of equations,'' \emph{Physical review letters}, vol. 103, no.~15, p. 150502,
  2009.

\bibitem{soeken18}
M.~Soeken, T.~Haener, and M.~Roetteler, ``Programming quantum computers using
  design automation,'' in \emph{Design, Automation \& Test in Europe Conference
  \& Exhibition (DATE), 2018}.\hskip 1em plus 0.5em minus 0.4em\relax IEEE,
  2018, pp. 137--146.

\bibitem{vad12}
V.~Kliuchnikov, D.~Maslov, and M.~Mosca, ``Fast and efficient exact synthesis
  of single qubit unitaries generated by {C}lifford and {T} gates,''
  \emph{arXiv preprint arXiv:1206.5236}, 2012.

\bibitem{amy19}
M.~Amy, P.~Azimzadeh, and M.~Mosca, ``On the controlled-{NOT} complexity of
  controlled-{NOT}–phase circuits,'' \emph{Quantum Science and Technology},
  vol.~4, no.~1, p. 015002, 2019.

\bibitem{meuli18}
G.~Meuli, M.~Soeken, and G.~De~Micheli, ``Sat-based $\{$CNOT, T$\}$ quantum
  circuit synthesis,'' in \emph{International Conference on Reversible
  Computation}.\hskip 1em plus 0.5em minus 0.4em\relax Springer, 2018, pp.
  175--188.

\bibitem{Nam18}
Y.~Nam, N.~J. Ross, Y.~Su, A.~M. Childs, and D.~Maslov, ``Automated
  optimization of large quantum circuits with continuous parameters,''
  \emph{npj Quantum Information}, vol.~4, no.~1, p.~23, 2018.

\bibitem{Bennett1989}
C.~H. Bennett, ``Time/space trade-offs for reversible computation,'' \emph{SIAM
  Journal on Computing}, vol.~18, no.~4, pp. 766--776, 1989.

\bibitem{Chan2015}
S.~M. Chan, M.~Lauria, J.~Nordstrom, and M.~Vinyals, ``Hardness of
  approximation in {PS}pace and separation results for pebble games,'' in
  \emph{2015 IEEE 56th Annual Symposium on Foundations of Computer Science
  (FOCS)}, vol.~00, 2015, pp. 466--485.

\bibitem{Knill1995}
E.~Knill, ``An analysis of {B}ennett's pebble game,'' \emph{arXiv preprint
  math/9508218}, 1995.

\bibitem{Komarath2018}
B.~Komarath, J.~Sarma, and S.~Sawlani, ``Pebbling meets coloring: reversible
  pebble game on trees,'' \emph{Journal of Computer and System Sciences},
  vol.~91, pp. 33--41, 2018.

\bibitem{nielsen00}
M.~A. Nielsen and I.~Chuang, ``Quantum computation and quantum information,''
  2000.

\bibitem{Nik08}
L.~de~Moura and N.~Bj{\o}rner, ``{Z3}: An efficient {SMT} solver,'' in
  \emph{Tools and Algorithms for the Construction and Analysis of Systems},
  C.~R. Ramakrishnan and J.~Rehof, Eds.\hskip 1em plus 0.5em minus 0.4em\relax
  Springer Berlin Heidelberg, 2008.

\bibitem{Costello2013}
J.~W. Bos, C.~Costello, H.~Hisil, and K.~Lauter, ``Fast cryptography in genus
  2,'' in \emph{Advances in Cryptology -- EUROCRYPT 2013}, T.~Johansson and
  P.~Q. Nguyen, Eds.\hskip 1em plus 0.5em minus 0.4em\relax Berlin, Heidelberg:
  Springer Berlin Heidelberg, 2013, pp. 194--210.

\bibitem{SRH18}
M.~Soeken, H.~Riener, W.~Haaswijk, and G.~De~Micheli, ``The {EPFL} logic
  synthesis libraries,'' \emph{arXiv preprint arXiv:1805.05121}, 2018.

\bibitem{sm17}
M.~Soeken, M.~Roetteler, N.~Wiebe, and G.~De~Micheli, ``Design automation and
  design space exploration for quantum computers,'' in \emph{2017 Design,
  Automation \& Test in Europe Conference \& Exhibition (DATE)}.\hskip 1em plus
  0.5em minus 0.4em\relax Ieee, 2017, pp. 470--475.

\bibitem{ibm17}
``{IBM} builds its most powerful universal quantum computing processors,''
  2017.

\bibitem{grover96}
L.~K. Grover, ``A fast quantum mechanical algorithm for database search,'' in
  \emph{Proceedings of the twenty-eighth annual ACM symposium on Theory of
  computing}.\hskip 1em plus 0.5em minus 0.4em\relax ACM, 1996, pp. 212--219.

\bibitem{barenco95}
A.~Barenco, C.~H. Bennett, R.~Cleve, D.~P. DiVincenzo, N.~Margolus, P.~Shor,
  T.~Sleator, J.~A. Smolin, and H.~Weinfurter, ``Elementary gates for quantum
  computation,'' \emph{Physical review A}, vol.~52, no.~5, p. 3457, 1995.

\bibitem{linke17}
N.~M. Linke, D.~Maslov, M.~Roetteler, S.~Debnath, C.~Figgatt, K.~A. Landsman,
  K.~Wright, and C.~Monroe, ``Experimental comparison of two quantum computing
  architectures,'' \emph{Proceedings of the National Academy of Sciences}, vol.
  114, no.~13, pp. 3305--3310, 2017.

\end{thebibliography}


\end{document}